\documentclass[12pt]{article}
\usepackage{epsf}

\usepackage{amsmath,amssymb}
\def\sl2R{sl(2,\mathbb{R})}
\def\SL2R{SL(2,\mathbb{R})}
\def\Real{\mathbb{R}}

\def\thefootnote{\fnsymbol{footnote}}
\def\11{\mbox{$1$}}

\renewcommand{\thefootnote}{\alph{footnote}}

\newcommand{\rref}[1]{(\ref{#1})}
\newcommand{\beqn}{\begin{equation}}
\newcommand{\eeqn}{\end{equation}}
\newcommand{\beqarr}{\begin{eqnarray}}
\newcommand{\eeqarr}{\end{eqnarray}}

\newcommand{\matc}{\begin{array}{c}}
\newcommand{\matcc}{\begin{array}{cc}}
\newcommand{\matccc}{\begin{array}{ccc}}
\newcommand{\matcccc}{\begin{array}{cccc}}
\newcommand{\emat}{\end{array}}

\newcommand{\IH}{\relax{\rm I\kern-.18em H}}
\newcommand{\IR}{\relax{\rm I\kern-.18em R}}
\newcommand{\IK}{\relax{\rm I\kern-.18em K}}
\newcommand{\II}{\hbox{\rm 1\kern-.28em I}}
\newcommand{\Is}{\relax{\rm 1\kern-.35em 1}}
\begin{document}

\begin{titlepage}

January 2002         \hfill
\vskip -0.55cm
\hfill  RU-02-1-B
\\
\begin{center}
\vskip .15in
\renewcommand{\thefootnote}{\fnsymbol{footnote}}
{\LARGE Quantum Mechanics on Noncommutative \\\vskip .15in Riemann Surfaces}
\vskip .25in
Bogdan Morariu$^{a}$  and
Alexios P. Polychronakos$^{b,}$\footnote{On leave from
Theoretical Physics Dept., Uppsala University, Sweden }
\vskip .25in

{\em    $^{a,\,b}$Department of Physics,        Rockefeller University   \\
        New York, NY 10021, USA     \\
\vskip .25in
        $^{b}$ Physics Department, University of Ioannina \\
        45110 Ioannina, Greece}
\end{center}
\vskip .2in
\vskip .05in {\hskip 1cm \rm E-mail: morariu@summit.rockefeller.edu, 
poly@teorfys.uu.se}
\vskip 0.25in
\begin{abstract}
We study the quantum mechanics of a charged particle on a constant 
curvature noncommutative Riemann surface in the presence of a constant
magnetic field. We formulate the problem by considering quantum mechanics
on the noncommutative $AdS_2$ covering space
and gauging a discrete  symmetry group which defines a genus-$g$ surface.
Although there is no magnetic field quantization on the covering
space, a quantization condition is required in order to have 
single-valued states on the Riemann surface. 
For noncommutative $AdS_2$ and sub-critical values of the magnetic field 
the spectrum has a discrete Landau level part as well as a continuum, 
while for over-critical values we obtain a purely noncommutative
phase consisting entirely of Landau levels.

\end{abstract}

\end{titlepage}

\newpage
\renewcommand{\thepage}{\arabic{page}}

\setcounter{page}{1}
\setcounter{footnote}{0}

\section{Introduction}
\label{Intro}
Noncommutative quantum field theories have been studied very intensely
over the last few years especially because of their relation to
M-theory compactifications~\cite{Connes:1998cr}
and string theory in nontrivial
backgrounds~\cite{Chu:1999qz,Schomerus:1999ug,Seiberg:1999vs}.
They are interesting because they preserve some of the nonlocal
properties inherent in string theory. For example, T-duality is a manifest
symmetry~\cite{Schwarz:1998qj,Brace:1999ku}.
(For recent  reviews of noncommutative gauge theory see~\cite{Har}.)
Recently, noncommutative Chern-Simons was shown to give an
alternative description of the fractional quantum Hall
effect 
\cite{Susskind,Polychronakos:2001mi,Hellerman:2001rj,
Karabali:2001xq,Morariu:2001qa}.

At low enough energies the single-particle sector becomes relevant and thus
it is enough to consider noncommutative quantum mechanics. (For  early
studies of noncommutativity in quantum mechanics see 
\cite{RJ,AFN,LSZ,BS,DH,GLR}.)
In particular, one can consider the quantum mechanics of
a charged particle moving on a two dimensional noncommutative surface in the
presence of a constant magnetic field. The problem on the plane and
the sphere has been considered
in~\cite{Nair:2000ct,Nair:2000ii,Karabali:2001te},
on the noncommutative torus in~\cite{Morariu:2001dv} and on 
noncommutative $AdS_2$ in \cite{Iengo:2001pi}.

In this paper we generalize this to higher
genus  noncommutative Riemann surfaces.
In Section~\ref{GaugeTheory} we review noncommutative $U(1)$ gauge theory on
$AdS_2$\,.  We study the quantum mechanics of a
charged particle on noncommutative Euclidean $AdS_2$
in a constant magnetic field in Section~\ref{QMechanics}.
This problem has also been considered in~\cite{Iengo:2001pi}; however,
because only representations of the Lie algebra
$\sl2R$ which integrate to representations of the group $\SL2R$\, were used,
a quantization of the magnetic field resulted. Such a
quantization is certainly
not observed in the commutative limit, since the topology of $AdS_2$
is trivial. We show that more general representations are allowed such
that the magnetic field is not quantized.

We also discuss the energy spectrum.
Unlike the usual Landau levels on the plane, for
commutative $AdS_2$ the Hamiltonian has both a
discrete spectrum and a  continuum. Semi-classically this can be
understood as follows: on a plane, for any finite energy the
classical orbits are closed and single valuedness of the wave function
phase around the orbit leads to a quantization of the energy.
On $AdS_2$, if the energy is above a threshold,
we have open trajectories and no quantization of the energy.
The spectrum for noncommutative $AdS_2$ is similar, except that
for a magnetic field above a critical value $B_{crit} = 1/\theta$ all
motion is bounded and there is only a discrete spectrum.

In Section~\ref{NCRS} we construct quantum mechanics on a noncommutative
Riemann surface by modding $AdS_2$ by a discrete
subgroup of $SO(2,1)$ which defines the cycles of a genus-$g$ surface.
Gauging of this discrete subgroup is just the
requirement that the Hilbert space is projected to states that
transform trivially
under the action of the subgroup, which corresponds to invariance of
(scalar) wavefunctions around the cycles, up to gauge transformations
and vacuum angles.
We show that this gauging
requires a certain quantization condition for the magnetic field and
demonstrate that in the commutative limit this condition reduces
to the standard Dirac quantization of the flux.
The Landau level spectrum for a noncommutative Riemann
surface is the same as that of $AdS_2$ but with finite degeneracy.
One also expects a discrete spectrum above the threshold, 
but little is known about this even in the commutative case.
A partial list of studies of the same problem on
a commutative Riemann surface
is~\cite{Comtet,Guztwiller}. The concept
of a noncommutative Riemann surface was also discussed
in~\cite{Bertoldi:2000hj}.

Finally, in the last section we briefly discuss some open issues for
future investigation.

\section{Gauge theory on the noncommutative $AdS_2$}
\label{GaugeTheory}

In this section we discuss $U(1)$ gauge theory on the
noncommutative $AdS_2$\,.
We follow closely the treatment of the noncommutative sphere
in~\cite{Karabali:2001te}. Field theory on the noncommutative sphere
was introduced in~\cite{Madore:1991bw} 
and studied rather extensively 
in~\cite{Carow-Watamura:1996wg}.
First consider the Lie algebra
\begin{equation}
[x_i,x_j]=i~\frac{\theta}{r} ~\epsilon_{ij}^{~~k} x_k~,\label{lie}
\end{equation}
where $\theta$ and $r$ are real parameters which we take to be
positive, $\epsilon_{123}=1$ and indexes are raised and lowered with the
metric $\eta={\rm diag}(1,1,-1)$.
The rescaled generators
$R_i=\frac{r}{\theta} x_i$ satisfy the $\sl2R$ relations
\begin{equation}
[R_i,R_j]=i~\epsilon_{ij}^{~~k} R_k~, \label{sl2r}
\end{equation}
with the quadratic Casimir
\begin{equation}
R^2 = R_1^2 + R_2^2 - R_3^2~.
\end{equation}

Let us briefly describe the unitary representations of $\sl2R$.
These representations are infinite dimensional since the metric
is of indefinite signature.
Usually, in the mathematical literature~\cite{Bargmann}
one finds the description
of the representations of the Lie algebra which can be integrated to true
representations of the groups
$\SL2R$ or $SO(2,1)$. While somewhat less familiar than the
unitary representations of $su(2)$, they can nevertheless be obtained
exactly in the same way.  One starts with an arbitrary $R_3$ eigenstate
$|m\rangle$ of unit norm and obtain other states in the representation
by applying $R_{\pm}=R_1\pm i R_2$. Using the fact that $R_i$ are
hermitian, one can calculate the norm of
these states and require it
to be positive. After this 
analysis~\cite{Dixon:1989cg,Plyushchay:1991hy,Maldacena:2000hw},
one obtains representations which
are of the following types:
\begin{itemize}
\item {\em Principal discrete series:}~These representations 
   act on the Hilbert space
   \begin{equation}
      {\cal D}_{j}^{\pm}=
      \{|j;m\rangle;~ 
      m=\pm j,\pm j \pm 1,\pm j \pm 2,\ldots \}~.\nonumber
   \end{equation}
   The state $|j;m\rangle$ has $R_3=m$, and the state $|j;-j\rangle$ has the 
   highest weight in ${\cal D}_j^{-}$ 
   while $|j;j\rangle$  has the lowest weight in ${\cal D}_j^{+}$\,.
   The Casimir equals $R^2=j(1-j)$ where $j$ is an arbitrary 
   positive {\em real} number. 
\item {\em Principal continuous series:}~These representations act 
  on the Hilbert space
  \begin{equation}
      {\cal C}_{j}^{\alpha}=
      \{|j,\alpha;m\rangle;~ 
      m=\alpha,\alpha\pm 1,\alpha\pm 2,\ldots \}~.\nonumber
  \end{equation}
  labeled by two continuous 
  parameters $j$ 
  and $\alpha$\,. The Casimir is given by 
  $R^2=j(1-j)$ for $j=1/2+ i s$ where $s$ is real and positive. The
  parameter $\alpha$ is real and can be chosen to satisfy $\alpha \in
[0,1)$. The states have  $R_3=m$.
\item {\em Complementary continuous series:}~These 
  representations act on the Hilbert space
  \begin{equation}
      {\cal E}_{j}^{\alpha}=
      \{|j,\alpha;m\rangle;~ 
      m=\alpha,\alpha\pm 1,\alpha\pm 2,\ldots \}~.\nonumber
  \end{equation}
  The parameter $\alpha$ is real and can be chosen to satisfy $\alpha \in
  [0,1)$\, while $j$ is real in the interval $j \in (1/2,1)$ and must satisfy 
  $j(1-j) > \alpha(1-\alpha)$\,.
\item {\em Identity representation:}~This is the trivial one
dimensional representation.
\end{itemize}

The representations in the discrete series form a discrete set
only if we require them
to integrate to representations of either the group
$\SL2R$ or $SO(2,1)$. Then,  $j$ must be an
integer or half integer for $\SL2R$,
while for  $SO(2,1)$ it must be an integer. In general, a unitary
representation of a semi-simple Lie algebra is also a unitary
representation of the universal covering group $\tilde{G}$
of all the groups $G$ with the
given algebra. Since such a  group $G$ has the form $G=\tilde{G}/\Gamma$
where $\Gamma$ is a discrete subgroup of  $\tilde{G}$\,,
to obtain representations of $G$ we must restrict to $\Gamma$-invariant
representations of $\tilde{G}$.
Equivalently, a necessary and sufficient condition for a representation
of a semi-simple Lie algebra to integrate to a representation of the
Lie group
$G$, is to be a good representation of a maximal compact subgroup of $G$.
Regarded as a Riemannian manifold
(with the metric given by the Killing metric), the
universal covering group of $\SL2R$ or $SO(2,1)$ is in
fact the familiar $AdS_3$ of unit radius and nonperiodic
time\footnote{For an illuminating discussion of $\SL2R$ and its covering
 group see ~\cite{BB}.}.
It has
the topology $D\times \Real$, where $D$ denotes a disk. We can obtain
$\SL2R$ by identifying time with period $4\pi$ and
$SO(2,1)$ by identifying time with period $2\pi$. Both groups have the
topology $D \times S^1$. This leads to the
quantization of $j$ described above.

Noncommutative $AdS_2$ of radius $r$ is defined as
the matrix algebra
generated by $x_i$ in the ${\cal D}^{+}_j$
irreducible unitary representation
where the Casimir satisfies
\begin{equation}
x^2 = x_1^2+x_2^2-x_3^2=-r^2~,\nonumber
\end{equation}
and $x_3$ is positive definite.
We must take $j>1$ so that $x^2$ be negative. Then the
parameter $\theta$ is given by
\begin{equation}
\theta=\frac{r^2}{\sqrt{j(j-1)}}~.\label{THETA}
\end{equation}
For states with $x_1 , x_2 \sim 0$, $x_3 \simeq r$, (\ref{lie}) reduces to
the planar noncommutativity relation $ [ x_1 , x_2 ] = -i \theta$ and thus
$\theta$ is identified as the noncommutativity parameter.
Note that for fixed $r$\,, since $j$ can vary continuously, there is
no quantization of $\theta$.

In the operator approach, scalar fields on noncommutative $AdS_2$ space
are defined as arbitrary operators on the Hilbert space and thus can be
identified with arbitrary elements of the algebra $\psi$.
We can implement the infinitesimal action of $\sl2R$ on the
generators of the noncommutative $AdS_2$ as
$[R_i,x_j]= i~\epsilon_{ij}^{~~k} x_k$. Since this action is a
derivation, we can define it also on an arbitrary element $\psi$ of the
algebra as
\begin{equation}
L_i(\psi)=[R_i,\psi]~.\nonumber
\end{equation}
We can then
define the derivative operators $\nabla_i=-\frac{i}{r}
\,R_i$ on $\psi$, which satisfy
\begin{equation}
[\nabla_i,\nabla_j]-\frac{1}{r}\epsilon^{~~k}_{ij} \nabla_k=0~.\nonumber
\end{equation}

We now formulate gauge theory on the noncommutative $AdS_2$.
The covariant derivative operators can be defined as a perturbation of
the derivative operators
\begin{equation}
D_i=\nabla_i +i A_i~.\nonumber
\end{equation}
Under gauge transformations, which are just time-dependent
infinite dimensional unitary matrices $U$, the covariant derivative
operators transform as
\begin{equation}
D'_i= UD_iU^{-1}~.
\end{equation}
It is convenient to also introduce covariant coordinates~\cite{Madore:2000en}
\begin{equation}
X_i = i\theta D_i = x_i -\theta A_i~,\nonumber
\end{equation}
parametrizing a noncommutative two-dimensional membrane.
The requirement that there be only two independent components of the
gauge field on $AdS_2$ is equivalent to the requirement that there be no
transversal excitations of the membrane. So the $X_i$ satisfy
the hyperboloid condition $X^2 = - r^2$, or, equivalently,
\begin{equation}
D^2= \left(\frac{r}{\theta}\right)^2 = \frac{j(j-1)}{r^2}~.\label{D2}
\end{equation}
This can be rewritten as
\begin{equation}
x^i A_i + A_i x^i -\theta A^2=0~.\label{trans}
\end{equation}
In the commutative limit $\theta \to 0$,~\rref{trans} is just the
condition that $A_i$
is tangent to the hyperboloid.

We can define a gauge covariant field strength as
\begin{equation}
iF_{ij}= [D_i,D_j]-\frac{1}{r} ~\epsilon_{ij}^{~~k}D_k~.\nonumber
\end{equation}
Notice that $F_{ij}=0$ for vanishing $A_i$ or any other gauge
equivalent configuration.
For a commutative time we also introduce $D_0=\partial_0 +i A_0$ and
define
\begin{equation}
iF_{0i}= [D_0,D_i]~.\nonumber
\end{equation}
Since the integral on $AdS_2$ is just
$\int \psi = 2\pi \theta \,{\rm Tr}(\psi)$
the Maxwell action takes the form
\begin{equation}
{\cal S} =
-\frac{1}{4g^2} \int dt \,2\pi \theta \,{\rm Tr}(F_{\mu \nu}F^{\mu \nu})~.
\nonumber
\end{equation}

\section{Quantum mechanics and spectrum on noncommutative $AdS_2$}
\label{QMechanics}

In this section we discuss the quantum mechanics of a charged
particle in a constant magnetic field on a noncommutative
$AdS_2$\,.

The magnetic field, defined as $B_i =
\frac{1}{2}\, \epsilon_{i}^{~jk} F_{jk}$\,, takes the form
\begin{equation}
iB_i = \epsilon_{i}^{~jk}D_j D_k +\frac{1}{r} D_i~.\label{Bfield}
\end{equation}
To have a uniform magnetic field we will take $B_i$
proportional to the gauge-covariant coordinate $X_i$
\begin{equation}
B_i = -\frac{B}{r} X_i = - \frac{i\theta B}{r} D_i~,\nonumber
\end{equation}
and this together with equation~\rref{Bfield} implies
\begin{equation}
[D_i, D_j]=\frac{1-\theta B}{r} \, \epsilon_{ij}^{~~k} D_k~,\nonumber
\end{equation}
which, up to a rescaling of $D_i$, are just the $\sl2R$
relations. Thus we have
\begin{equation}
D_i =-i\,\frac{1-\theta B}{r} K_i~,
\end{equation}
where $K_i$
satisfy the algebra~\rref{sl2r}. Since $D_i$ still have to satisfy (\ref{D2}),
we take the representation of
$K_i$ to be irreducible and of the form ${\cal D}^{\pm}_s$ with $s>1$. 
We will show shortly that the choice of ${\cal D}^{+}_s$ or
${\cal D}^{-}_s$ depends on the value of $B$\,.
By a gauge transformation we can bring the $K_i$ in
the standard form where $K_3$ is diagonal.
The relation~\rref{D2} implies that $s$ must satisfy
\begin{equation}
(1-\theta B)^2=
\frac{j(j-1)}{s(s-1)}~.\label{NoQ}
\end{equation}
Since neither $j$ nor $s$ are quantized when considering unitary
representations of the Lie algebra, the relation~\rref{NoQ} does not
imply any quantization of $B$ as was assumed in~\cite{Iengo:2001pi}.
This result is compatible with the commutative
limit where $B$ is not quantized, since $AdS_2$ has a trivial topology.

For a charged field $\psi$, with the gauge transformation
$\psi'=U \psi$\,, we define the covariant derivative action as
\begin{equation}
D_i(\psi)=
D_i \psi -\psi \nabla_i~.\nonumber
\end{equation}
On the right hand side, $D_i$ represents an element of the algebra while
on the left hand side it denotes an action on $\psi$. We can also write this as
\begin{equation}
i D_i(\psi)=
\frac{1}{r}(\gamma K_i  \psi -\psi R_i)~,\label{DI}
\end{equation}
where $\gamma =1-\theta B$\,.

Note that $\psi$ is a matrix multiplied on the left by ${\cal D}_s^{\pm}$
representation matrices and on the right by ${\cal D}_j^+$
representation matrices. It is more
convenient to have both of these multiplications described as
actions on the left. Since the generators are
hermitian, transposition is equivalent to complex conjugation and this
takes ${\cal D}_j^+$ into ${\cal D}_j^-$.
Concretely, to the matrix $\psi_{nm}$ we associate the state
\begin{equation}
|\psi\rangle=
\sum_{n,m=0}^{\infty}
\psi_{mn}
|s+m\rangle_s^{\pm}
|-j-n\rangle_j^-~,\nonumber
\end{equation}
and then the relation~\rref{DI} can be written as
\begin{equation}
i D_i|\psi\rangle=
\frac{1}{r}(\gamma {\cal R}_{i}^{(s)\pm} + {\cal
  R}_{i}^{(j)-})|\psi\rangle~,
\nonumber
\end{equation}
where ${\cal R}_{i}^{(s)\pm}$\, (${\cal R}_{i}^{(j)-}$)\, denote
operators acting on states
of the ${\cal D}_s^{\pm}$\, (${\cal D}_j^{-}$\,) representations.
In this notation, the action
of the generators  $J_i$ of the $\sl2R$\,, representing the infinitesimal
symmetry of $AdS_2$\,,
takes the form
\begin{equation}
J_i|\psi\rangle=
( {\cal R}_{i}^{(s)\pm} + {\cal R}_{i}^{(j)-})|\psi\rangle~.\label{Inf}
\end{equation}
In particular, $J_3$ can be identified with angular momentum around
the origin.

The equation of motion for $\psi$ can be obtained from an action of
the Schr\"{o}edinger type
\begin{equation}
{\cal S}=\int dt\,
2\pi \theta \,{\rm Tr} \left(i\psi^{\dagger}\dot{\psi} +
\frac{1}{2} D_i(\psi)^{\dagger}  D^i(\psi)\right)~.\label{Sch}
\end{equation}
Then the Hamiltonian is given by $H=-\frac{1}{2} D^2$\,, and with a
little bit of algebra it can be rewritten as
\begin{equation}
H=\frac{\gamma}{2r^2}\left(J^2+ \left(\frac{Br^2}{\gamma}\right)^{2}\right)~.
\label{Ham}
\end{equation}
The spectrum and eigenstates of the Hamiltonian are trivially related
to those of $J^2$, and thus they are given by pure representation
theory. They can be obtained from the following 
tensor product decompositions
\begin{eqnarray}
{\cal D}_s^{-}\otimes {\cal D}_j^{-} &=&
\sum_{m=0}^{\infty} {\cal D}_{s+j+m}^{-}~,\label{DMM}\\
{\cal D}_s^{+}\otimes {\cal D}_j^{-} &=&
\sum_{n \in I} {\cal D}_{\alpha+n}^\pm \oplus
\int_{s= 0}^{\infty} {\cal C}_{1/2+i s}^{\alpha}
\oplus \ldots
~,\label{DPM}  
\end{eqnarray}
where $I=\{n \in \mathbb{Z}; 1/2<\alpha+n\leq
|s-j|\}$\,. In~\rref{DPM}
the $+$ sign is taken for $s>j$\,
and  $\alpha=|s-j|{\rm mod}(1)
\in [0,1)$.
Note that in~\rref{DPM}, the rhs.~contains
representations from both the discrete and the continuous series,
and that the discrete series start at $k \ge \frac{1}{2}$.
The dots in~\rref{DPM} stand for complementary series
representations. However, in the expansion of a normalizable state in terms of
energy eigenstates the complementary series representations
have zero measure~\cite{BB}, thus they do not contribute to the
spectrum.   

To choose between ${\cal D}^{+}_s$ and ${\cal D}^{-}_s$\,
we require that the Hamiltonian~\rref{Ham} be bounded from below.
For $B<1/\theta$\,, since $\gamma$\, is positive we choose  ${\cal
D}^{+}_s$. By~\rref{DPM} there is only a finite number of
discrete series representations and because of the second term
in~\rref{Ham} the Hamiltonian is positive definite.
The spectrum consists of a finite set of discrete Landau level energies
\begin{equation}
E_n=\frac{\gamma}{2r^2}\left((\alpha+n)(1-\alpha-n)+
\left(\frac{Br^2}{\gamma}\right)^{2}\right)~,~n \in I~,
\nonumber
\end{equation}
above which there is a continuous spectrum, starting at the threshold
energy
\begin{equation}
E_{thres} =\frac{\gamma}{8r^2} + \frac{B^2 r^2}{2\gamma}~.
\label{thres}
\end{equation}

For $B>1/\theta$\,, since $\gamma$\, is negative, we
have a Hamiltonian bounded from below if we choose  ${\cal
D}^{-}_s$\,. In this case there is only a discrete energy
spectrum given by
\begin{equation}
E_n=\frac{\gamma}{2r^2}\left((j+s+n)(1-j-s-n)+
\left(\frac{Br^2}{\gamma}\right)^{2}\right)~,~n=0,\ldots ,\infty~.
\nonumber
\end{equation}
This phase is a purely noncommutative one.

We can check that, in the limit $r^2 \to \infty$ with constant $\theta$,
the above spectrum reproduces the Landau levels on the
noncommutative plane found in \cite{Nair:2000ii}.  In that limit the
continuous spectrum is pushed to infinity. For the discrete levels
we have, up to $O(r^{-2})$ corrections,
\begin{equation}
j = \frac{r^2}{\theta} + \frac{1}{2}  ~,~~
s = \frac{r^2}{|\gamma| \theta} + \frac{1}{2} ~,~~
E_n=(n+ \frac{1}{2}) |B| ~,
\nonumber
\end{equation}
in agreement with the planar result. The density of states agrees as well.
This gives an independent justification for the choice of ${\cal D}^{-}_s$\,
for the representation of the covariant derivatives in the case $B>1/\theta$,
since the system maps to the correct over-critical planar phase.
At the commutative $AdS_2$ limit, $\theta \to 0$ for constant $r$,
we recover the standard results \cite{Comtet}.

\begin{figure}[htb]
\begin{center}
\epsfxsize=4.0in\leavevmode\epsfbox{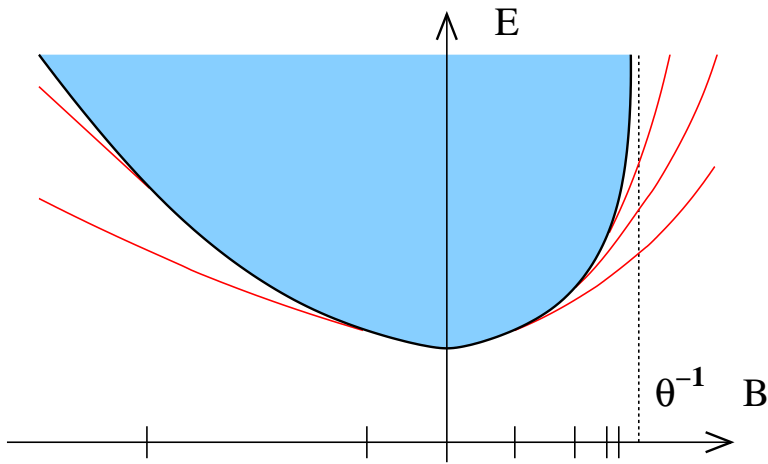}
\end{center}
\caption{Continuous  spectrum and Landau levels.}
\label{fig1}
\end{figure}
The form of the spectrum we obtained is depicted in Figure \ref{fig1}:
for small positive values of the magnetic field ($|s-j|<\frac{1}{2}$) 
the spectrum is entirely continuous,
with a threshold as in (\ref{thres}). For positive $B$ corresponding to $s-j = 
\frac{1}{2}$, a single Landau level `peels' from the bottom of the continuum.
For $s-j = \frac{3}{2}$ a second level peels, and so on. As $B \to B_{crit} = 
1/\theta$, an infinity of Landau levels has formed, while the continuum is
pushed to infinity. Above $B_{crit}$ no more Landau levels are formed and
there is no continuum. Similarly, for negative values of $B$, Landau levels
peel from the continuum at points at which $s-j$ equals negative half-integers.
Since $s>1$, there is a lowest such point, 
for $s-j = -[j-\frac{3}{2}] - \frac{1}{2}$
corresponding to some $B_\ell$, at which the last Landau level peels. 
For $B < B_\ell$ no more Landau levels form. We see that, for negative $B$,
there is a maximum number of Landau levels $N_{max} = [j - \frac{3}{2}]$.
For large $r$ or small $\theta$, $N_{max} \sim
\frac{r^2}{\theta}$. The entirely discrete spectrum 
above $B_{crit} = 1/\theta$ and the existence of $N_{max}$
are purely noncommutative effects.

\section{Noncommutative Riemann surfaces}
\label{NCRS}
In this section we will formulate quantum mechanics on a
noncommutative Riemann surface by gauging a discrete symmetry group of
the action~\rref{Sch}.
To set the stage, we first review how to obtain a
commutative Riemann surface endowed with a constant curvature metric
by modding out the upper half-plane (or the
mass hyperboloid) by the action of a Fuchsian group~\cite{Farkas}.

Consider a
Riemann surface $\Sigma$ of genus $g$ on which we have chosen a canonical
homology basis with generators $a_i,~b_i,~i=1,\ldots ,g$\,, i.e. the
intersection numbers are  given by
\begin{equation}
a_i \wedge a_j= 0~,~
a_i \wedge b_j= \delta_{ij}~,~b_i \wedge b_j= 0~.
\end{equation}
Let us pick a
representative in the homology class of each generator which also goes through
a fixed point $P$ on $\Sigma$. Then, $a_i$\, and $b_i$\, can be
interpreted as generators of the fundamental group $\pi_1(\Sigma)$
based at $P$ of the surface $\Sigma$.
As such, they satisfy
\begin{equation}
\prod_{i=1}^{g} \left(
a_i b_i a_i^{-1} b_i^{-1}
\right)
=1~.\label{cut}
\end{equation}
To understand the equation~\rref{cut}, take the above homology generators
passing through  $P$ to be geodesics and then cut $\Sigma$ along
them. The resulting surface, called the cut Riemann surface
$\Sigma_c$\,, is a $4g$-gon and the product on the
lhs. of~\rref{cut} is just the boundary cycle. This is obviously
contractable to a point.

The group of isometries of the mass hyperboloid $x^2= -r^2$ is $SO(2,1)$.
Group elements of $SO(2,1)$ acting without a fixed point are
called hyperbolic (they are
called elliptic if they have a finite fixed point and parabolic if the
fixed point is at infinity). Consider a discrete subgroup $\Gamma$
of $SO(2,1)$
isomorphic to the fundamental group $\pi_1(\Sigma)$ and
containing only hyperbolic elements. Then
$\Gamma$ must be generated by $g_{a_i}$ and $g_{b_i}$ satisfying
\begin{equation}
\prod_{i=1}^{g} \left(
g_{a_i}^{~} g_{b_i}^{~} g_{a_i}^{-1} g_{b_i}^{-1}
\right)
=1~. \label{Fuchs}
\end{equation}
All the nondegenerate Riemann surfaces of genus $g$\,
can be obtained by modding out the
mass hyperboloid by the action of such a group $\Gamma$\,. One can chose a
covering of the hyperboloid such that each fundamental region is
isomorphic to the cut Riemann surface $\Sigma_c$\,.

The action~\rref{Sch} is invariant under the infinitesimal $\sl2R$
transformations~\rref{Inf}. These transformations, actually, involve both
space translations and gauge transformations. They commute with the
Hamiltonian and correspond to the magnetic translations of the particle.
They are, thus, the appropriate transformations to be used in order to reduce
the Hilbert space to the one of the genus-$g$ Riemann surface.
As we will see, if $j$ and $s$ are chosen
appropriately, one can integrate the infinitesimal action of these generators
and  represent the group $\Gamma$ on the set of states.

We therefore define quantum mechanics on
the noncommutative Riemann surface as the system obtained by gauging the
group $\Gamma$, in analogy to the commutative case.
Since this group is discrete this just means that we
must project onto the subspace of gauge invariant states. More generally,
we can require invariance up to a phase (vacuum angle)
\begin{equation}
U(g_{\alpha})
\,\psi\,
V^{-1}(g_{\alpha})
=e^{i \xi_{\alpha}}\,\psi ~,~\alpha = 1,\ldots ,2g~.\label{Mod}
\end{equation}
In the above, the index $\alpha$ runs over the
$a_i$ and $b_i$ cycles, while $U$\, and $V$\, denote the
${\cal  D}_s^{\pm}$ and ${\cal D}_j^{+}$ representations of $g_{\alpha}$\,.
For $s=j$ the set of $\psi$'s  satisfying~\rref{Mod} form the 
algebra of ``functions'' on the noncommutative Riemann surface.
For $s\ne j$ the set of $\psi$'s satisfying~\rref{Mod} define a
projective module which is the noncommutative analogue of the set of
sections of a vector bundle.

Using~\rref{Mod} repeatedly we obtain the consistency condition
\begin{equation}
U(\prod_{i=1}^{g} \left(
g_{a_i}^{~} g_{b_i}^{~} g_{a_i}^{-1} g_{b_i}^{-1}
\right))
~\psi~
V^{-1}(\prod_{i=1}^{g} \left(
g_{a_i}^{~} g_{b_i}^{~} g_{a_i}^{-1} g_{b_i}^{-1}
\right))
=\psi ~. \label{cocycle}
\end{equation}
As we will now show, equation~\rref{cocycle} implies a quantization
of $\pm s-j$\,.

For $j$\, and $s$\, integers, ${\cal  D}_s^{\pm}$ and ${\cal D}_j^{+}$
are also representations of $SO(2,1)$\, and the
relation~\rref{Fuchs} implies
\begin{equation}
U(\prod_{i=1}^{g} \left(
g_{a_i}^{~} g_{b_i}^{~} g_{a_i}^{-1} g_{b_i}^{-1}
\right)) =
V(\prod_{i=1}^{g} \left(
g_{a_i}^{~} g_{b_i}^{~} g_{a_i}^{-1} g_{b_i}^{-1}
\right)) =1~,
\end{equation}
thus the consistency condition~\rref{cocycle} is satisfied
trivially.  However, for $j$\, and $s$\, {\em real}
positive,  since the representations $U$ and $V$ are only
representations of the universal covering group $S\widetilde{O(2,}1)$
we only have
\begin{equation}
U(\prod_{i=1}^{g} \left(
g_{a_i}^{~} g_{b_i}^{~} g_{a_i}^{-1} g_{b_i}^{-1}
\right)) =e^{i\Theta_s^{\pm}}~,~
V(\prod_{i=1}^{g} \left(
g_{a_i}^{~} g_{b_i}^{~} g_{a_i}^{-1} g_{b_i}^{-1}
\right)) =e^{i\Theta_j}~,\label{Phase}
\end{equation}
as we will explain shortly. In this case, the consistency
condition~\rref{cocycle}
is satisfied if the two phases in~\rref{Phase} are equal.
The origin of the above phases is as follows: Since all the $g_{\alpha}$ are
hyperbolic, they can be written as exponentials of elements in the Lie algebra.
Using the exponential map, $g_{\alpha}$ can also be understood as
group elements in the universal covering group $S\widetilde{O(2,}1)$\,. The
product on the lhs. of~\rref{Fuchs} with the multiplication performed
in the universal covering group does not necessarily give the identity
but some element of  $S\widetilde{O(2,}1)$\,
which projects to the identity of $SO(2,1)$\,. By looking at the form
of the $R_3$\, (or $K_3$\,) generators one can see that such an
element is represented by a phase.

Let us associate to each $g_{\alpha}$ a curve in $SO(2,1)$ denoted
$g_{\alpha}(t)$ representing a portion of a one dimensional subgroup
passing through  $g_{\alpha}$ such that
$g_{\alpha}(0)$ is the identity and $g_{\alpha}(1)=g_{\alpha}$.
Then to the product $\prod_{i=1}^{g} \left(
g_{a_i}^{~} g_{b_i}^{~} g_{a_i}^{-1} g_{b_i}^{-1}
\right)$\, we associate a curve of length $4g$ by translating
and joining the curves $g_{\alpha}(t)$\, in the obvious way: 
for $t\in [0,1)$ the
curve is given by $g_{a_1}(t)$; for $t\in [1,2)$ the
curve is given by $g_{a_1}(1)g_{b_1}(t-1)$\, ; and so on. Due to the
relation~\rref{Fuchs} this must be a closed curve in
$SO(2,1)$. However, the curve winds $2(g-1)$
around the noncontractable $S^1$ cycle of $SO(2,1)$\, and thus it is
an open curve in  $S\widetilde{O(2,}1)$\,.

Before we calculate the winding in our problem,
let us describe one way of obtaining it for an arbitrary closed
curve $g(t)$  in $SO(2,1)$. Fix a reference
point $P$ on the hyperboloid and a reference tangent vector at $P$\,.
The curve $g(t)P$\, is a closed curve on the hyperboloid.
The action of  $g(t)$\, on the reference vector gives a periodic vector
field around the curve $g(t)P$\,. The winding is just the number of times the
vector spins around itself as it goes once around the curve and
is a topological invariant.

In our problem the curve $g(t)P$\, is
just the boundary of $\Sigma_c$\,, and the reference vector is
parallel transported around the boundary of $\Sigma_c$\,.
Under parallel transport on a
hyperboloid of radius $r$\, around a closed loop enclosing an area $A$,
a vector  is rotated by an angle $\phi = A/r^2$. Since the scalar
curvature is given by $R= -2/r^2$\,, using the Gauss-Bonnet theorem
one can find the area of the surface $\Sigma$ to be
$A=4\pi(g-1)r^2$\,. Thus under parallel transport around  $\Sigma$ a
vector rotates an angle $\phi =4\pi(g-1)$\,.
Since we have $e^{i R_3 \phi}=e^{i \Theta_j}$\,,  the phase is given by
$e^{i\Theta_j}=e^{4\pi i (g-1)j}$\,.
The group $\Gamma$\, defined by the relation~\rref{Fuchs} is only
represented projectively
\begin{equation}
V(\prod_{i=1}^{g} \left(
g_{a_i}^{~} g_{b_i}^{~} g_{a_i}^{-1} g_{b_i}^{-1}
\right)) =e^{4\pi i(g-1)j}~.
\end{equation}
Projective representations of $\Gamma$\, were also considered
in~\cite{Bertoldi:2000hj} where they were obtained with the help
of a gauge field on
the Poincare plane. Here we see that projective representations
naturally occur if $j$\, is not an integer or half-integer.
Finally, the consistency
condition~\rref{cocycle} implies the quantization
\begin{equation}
\pm s-j=
\frac{n}{2(g-1)}~,\label{SJ}
\end{equation}
where $n$ is an arbitrary integer.

{}From experience with the noncommutative sphere and torus we
know that a more relevant quantity is a rescaled magnetic field
$\widetilde{B} \equiv B(1-\theta B)^{-1}$\,. 
This would be the strength of the
Seiberg-Witten mapped commutative gauge field in the planar case.
{}From equation~\rref{NoQ} we obtain
\begin{equation}
\widetilde{B}
=
\frac{1}{r^2}\left(
\pm \sqrt{s(s-1)}-\sqrt{j(j-1)}
\right)~.\label{BTquant}
\end{equation}
Since  $j$\, is fixed for a given $r$\, and $\theta$\, by
relation~\rref{THETA}, and $\pm s-j$ is quantized as in~\rref{SJ}
we see that $\widetilde{B}$ can only take discrete values. Note
however that, unlike the commutative case, the values of $\widetilde{B}$
are not equally spaced.

As a check, consider the
commutative limit, obtained by taking $j$ and $s$ to infinity while
keeping $r$ and $B$ finite (we must choose ${\cal D}_s^+$).
In this limit we have
$B =\frac{1}{r^2}(s-j)$\,, thus we must keep $s-j$
finite. Using this, we obtain the following integral quantization for
the flux
\begin{equation}
\Phi \equiv
A B =2\pi n~.
\end{equation}
This is the expected Dirac quantization (or integrality of the first
Chern number).

\section{Concluding remarks}
\label{LLL}

We have formulated the problem of a charged particle
on a noncommutative genus-$g$ Riemann surface and found the
condition required for the existence of scalar wavefunctions. The
spectrum of the particle, on the other hand, has not been fully identified.
To achieve this, we would need to identify the physical states which
satisfy the genus-$g$ condition (\ref{Mod}). This is, in principle, a purely
group-theoretic problem. We expect the degeneracy of each discrete Landau
level to become finite, and also the continuous spectrum to be fragmented
into discrete nondegenerate states. Besides the Landau levels, there
should be additional states below the threshold corresponding to
linear combinations of complementary series states. It is interesting
to understand how the continuous spectrum for the 
noncommutative $AdS_2$ emerges 
from the discrete spectrum of the noncommutative Riemann surface as
we take the genus $g$ and thus the area to infinity. In particular
note that only complementary series states contribute to the
continuum. As we take the area to infinity the density of states
increases, but to obtain a continuum the density must scale  
as the square root of the area. Presumably, this is what happens for
the principal but not for the complementary continuous representation states.
Carrying out this calculation and
identifying the full spectrum and degeneracies is a very interesting open
issue.

The Dirac-like quantization condition for the strength of the magnetic field
was derived by demanding invariance of the wavefunction under
magnetic translations around the cycles of the noncommutative Riemann surface.
It should be stressed that, as in the noncommutative torus case and unlike the
sphere, this is not a requirement for consistency of the problem.
In fact, we could have promoted the wavefunction $\psi$ into a multicomponent
vector by tensoring it with an $N$-dimensional vector space $V_{_N}$ and demand
invariance under combined magnetic translations and $U(N)$ transformations,
which would have resulted in an $N$-fold decrease in the unit of quantization
in (\ref{SJ}). This corresponds to `overlapping' $N$ copies of the fundamental
domain of the Riemann surface.

In the toroidal case~\cite{Morariu:2001dv},
the problem can be analyzed entirely in the canonical
framework by defining physical coordinate and momentum variables which
are well-defined on the torus. The representation theory of the
algebra of these
observables reproduces the above extended wavefunctions. In the genus-$g$
case there is no immediately obvious complete set of such observables.
Formulating and analyzing the noncommutative Riemann
problem in terms of such canonical observables is an interesting open problem.

Finally, we should remark that, although here we have only considered
$AdS_2 = \SL2R / \,U(1)$\,, it is obvious that the construction can be
generalized to $G/H$ where $G$ is a real semisimple Lie group and
$H$ is its maximal compact subgroup. Application of this technique to
physically relevant situations, such as the noncommutative gravity
setting of \cite{Nair:2001kr},
would be an interesting possibility. Moreover, the
methods developed in this paper could also 
be applied to the study of D-branes on
$AdS_3$\,, see \cite{Bachas:2000fr} and references therein.

\section*{Acknowledgments}
We would like to thank K. Bering for discussions and C. Bachas for
pointing out the connection to D-branes.
This work was supported in part by the~ U.S.~ Department~ of~ Energy~
under ~Contract Number DE-FG02-91ER40651-TASK B.

\end{document}